\documentclass{aa}
\usepackage{txfonts}
\usepackage{natbib}
\bibpunct{(}{)}{;}{a}{}{,}    
\usepackage{graphicx}
\usepackage{subfigure}
\usepackage{placeins}
\usepackage{float}
\usepackage{amsmath}
\usepackage{enumitem}
\usepackage[breaklinks=true, colorlinks=true, urlcolor=magenta, citecolor=blue, linkcolor=blue]{hyperref}

\begin{document}

\title{A tale of three cataclysmic variables with distinct superhumps}
\author{
        Arti Joshi\inst{\ref{PUC} 
   \thanks{E-mail: ajoshi@astro.puc.cl, 
 aartijoshiphysics@gmail.com}},
    Claus Tappert$^{2}$, 
    Márcio Catelan$^{1,3}$, 
    Linda Schmidtobreick$^{4}$,
    Mridweeka Singh$^{5}$
            }
\institute{
Institute of Astrophysics, Pontificia Universidad Católica de Chile, Av. Vicuña Mackenna 4860, 7820436 Macul, Santiago, Chile \label{PUC}
\and
Instituto de F\'isica y Astronom\'ia, Universidad de Valpara\'iso, Valpara\'iso, Chile
\and
Millennium Institute of Astrophysics, Nuncio Monse\~{n}or S\'{o}tero Sanz 100, Providencia, Santiago, Chile \label{MIA}
\and
European Southern Observatory, Casilla 19001, Santiago 19, Chile
\label{IIA}
\and
Indian Institute of Astrophysics, Koramangala, Bangalore 560034, India
}


\abstract{%
Superhumps are one of the most commonly observed variable features in the light curves of cataclysmic variables (CVs). To study the superhump behaviour of CVs, we present the Transiting Exoplanet Survey Satellite (TESS) observations of three CVs, namely CRTS J110014.7+131552, SDSS J093537.46+161950.8, and [PK2008] HalphaJ130559. Among them, a superoutburst is observed in CRTS J110014.7+131552 which is associated with the precursor outburst, where prominent superhumps are observed during maximum of the outburst with a mean period of  0.06786(1) d. We have observed variations in the superhump period, along with changes in the shape of the light curve profile and the amplitude of the superhumps during different phases of the outburst, indicating disc-radius variation as well as periodically variable dissipation at the accretion stream's bright spot. The data on SDSS J093537.46+161950.8 reveal previously unknown variations modulated with periods  
  0.06584(2) d and 2.36(2) d, related to the positive superhump and the disc-precession periods, respectively, which can reasonably be interpreted as a result of the prograde rotation of an eccentric accretion disc. Despite its short orbital period, the lack of outburst activity, along with its stable long-term brightness, discovery spectrum, and absolute magnitude suggests that the object might not be an SU UMa type dwarf nova. Instead, it may belong to the group of high mass-transfer CVs below the period gap, either to a rare class of nova-like variables or to the class of high-luminosity intermediate polars, a subclass of magnetic CVs. For [PK2008] HalphaJ130559, a new average orbital period of 0.15092(1) d has been identified. Additionally, this system displays previously undetected average periods of  0.14517(3) d and 3.83(1) d, which can be provisionally identified as negative superhump and disc-precession periods, respectively. If the identified simultaneous signals do indeed reflect negative superhump and disc-precession period variations then their origin may be associated with the retrograde precession of a tilted disc and its interaction with the secondary stream.
}

\titlerunning{CRTS J110014.7+131552, SDSS
J093537.46+161950.8, and [PK2008] HalphaJ130559}
\authorrunning{Joshi et~al.}
\maketitle


\section{Introduction}
\label{sec:intro}
Cataclysmic Variables (CVs) are close interacting binaries consisting of a low-mass main-sequence star transferring matter to the white dwarf (WD) via Roche lobe overflow \citep[see][for an exhaustive overview]{Warner95}. In non-magnetic or weakly magnetic systems, the accreting material forms an accretion disc around the WD due to conservation of the angular momentum, where viscous forces gradually drive it inward until it settles on the surface of the WD. Among non-magnetic CVs, nova-like variables form a 
  subclass defined by their non-eruptive behavior, as they have never shown nova or dwarf nova outbursts. Their high mass-transfer rates create ionized accretion discs, which prevent the disc-instability mechanism responsible for outbursts \citep{Osaki74}. On the other hand, recurring outbursts on timescales from days to several years often occur in the thermally unstable accretion discs of dwarf novae, depending on the mass-transfer rate. Among dwarf novae, the subtype called SU UMa lies below the period gap of 2-3 h and exhibits super-outbursts in addition to the normal outbursts. The thermal-tidal instability model (TTI), which integrates both thermal and tidal instabilities, is generally accepted as a valid explanation for superoutbursts and supercycle events \citep{Osaki89, Osaki96}. 
 During superoutbursts, SU UMa systems exhibit superhumps, which reflect the beat period between the orbital period and the disc precession period \citep{Vogt74, Osaki96, Wood11}. There are two types of superhumps: positive and negative ones, depending on whether the difference between the superhump and the orbital period is positive or negative, respectively. Positive superhumps are commonly found in the SU UMa-type systems. The period and amplitude of the positive superhumps fluctuate throughout the superoutburst \citep{Kato09}. The theory of negative superhumps is less developed, and research suggests that they result from retrograde precession of the tilted disc nodal line \citep{Bonnet-Bidaud85, Harvey95}, but there is still no consensus on the cause of the tilted disc. 
  
\begin{table*}
\normalsize
\caption{Log of the TESS observations of all three sources, J1100, J0935, and J130559.\label{tab:obslog}}
\setlength{\tabcolsep}{0.1in}
\centering
\begin{tabular}{@{}ccccccccc@{}}
\hline
Object&Sector&TIC IDs&Start Time (UTC) & End Time (UTC) & Total observing days     \\
\hline                    
J1100    & 72 & 903453082& 2023-11-11T16:19:23.8    & 2023-12-07T01:48:57.9 & 25.4\\
J0935    & 72 &840126870& 2023-11-11T16:22:10.6    & 2023-12-07T02:01:44.5 & 25.4\\
J130559  & 64 &253410027& 2023-04-06T14:47:49.9    & 2023-05-03T12:50:28.6 & 26.9\\  
         & 65 &253410027& 2023-05-04T05:48:28.8    & 2023-06-01T03:09:51.1 & 27.8\\
\hline
\end{tabular}
\end{table*}

To effectively detect superoutbursts, superhumps, and their variations, long-span data are essential. Therefore, the continuous, long-baseline, high-cadence optical photometric data from the Transiting Exoplanet Survey Satellite \citep[TESS;][]{Ricker15} provide an excellent resource for studying the detection and variation of superhumps. In this paper, we select a sample of three very interesting but insufficiently studied candidates to carry out a detailed study of these phenomena. The available information on these sources is summarized below.

CRTS J110014.7+131552 (hereafter J1100) was identified as a CV based on the
Sloan Digital Sky Survey (SDSS) spectrum \citep{Szkody06}. Later, \cite{Kato09} reported that superhumps were detected during the 2009 outburst observed by the Catalina Real Time Transient Survey \citep[CRTS\footnote{\url{http://nesssi.cacr.caltech.edu/DataRelease/}};][]{Drake09}. Using the short baseline of their observations, they determined a period of 0.067569 d.

SDSS J093537.46+161950.8 (hereafter J0935) was classified as CV by \cite{Szkody09} using the SDSS spectrum. The strong He II ($\lambda$ 4686 \AA) emission led \cite{Szkody09} to propose that it may harbor a magnetic WD or possibly be an old nova. \cite{Southworth15} reported photometric observations of J0935, revealing eclipses and allowing the determination of an orbital period  corresponding to 0.0640591 d. They were unable to definitively classify or understand the nature of this system based on their findings. Later, \cite{Hardy17} presented a single eclipse light curve at high time resolution using ULTRACAM on the William Herschel Telescope and suggested that its light curve resembles that of the polar HU Aqr.

 [PK2008] HalphaJ130559 (hereafter J130559) was selected as a CV based on the detection of H$\alpha$ emission \citep{Pretorius08}. Their spectroscopic investigation also revealed the presence of a strong He II ($\lambda$4686 \AA) emission line, together with broad Balmer and He I emission lines. Based on time-resolved spectroscopic observations, a period of $3.928\pm0.013$~h was determined and assumed to be the orbital period of the system. However, no photometric observations were conducted to further investigate this system.

Subsequent to these initial investigations, no further studies have been conducted on these systems. The inadequate examination of these sources prompted us to undertake a comprehensive investigation into their nature. Consequently, our current work focuses on studying  these systems utilizing extensive TESS observations. The paper is organised as follows. Sect. \ref{sec:obs} contains observations and data reduction. The analyses and results are discussed in Sect. \ref{sec:analysis}. Finally, the discussion and summary are presented in Sects. \ref{sec:diss} and \ref{sec:sum}, respectively.

\begin{figure}
\centering
\hspace*{-0.5cm}
\subfigure[]
{\includegraphics[width=0.5\textwidth]{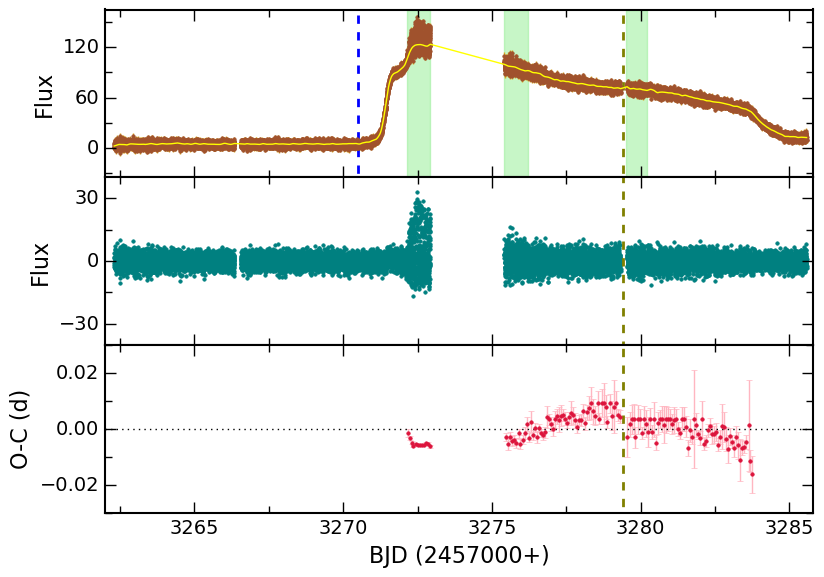}\label{fig:tesslc_J1100}}
\subfigure[]
{\includegraphics[width=0.5\textwidth]{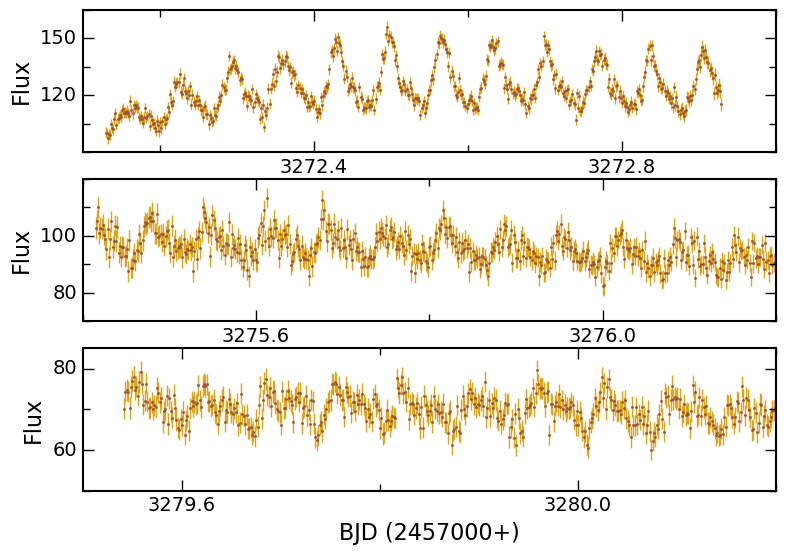}\label{fig:tesszoomlc_J1100}}
\caption{(a) TESS light curve of J1100, displaying a superoutburst. A vertical blue dashed line at BJD 2460270.5020 represents a conservative estimate of the start time of the outburst. The solid yellow line represents the smoothed light curve obtained using the LOESS fit. Middle: the detrended light curves after subtracting the smoothed light curve. Bottom: O-C curves of superhumps. The light green vertical dashed lines represent a clear period transition in the O-C trends.  (b) Zoomed-in segment of the TESS light curve over $\sim$ 0.7-0.8 d, corresponding to the shaded regions highlighted in the top panel of Figure \ref{fig:tesslc_J1100}, shown for clarity.} 
\end{figure}

\begin{figure}
\centering
\includegraphics[width=0.48\textwidth]{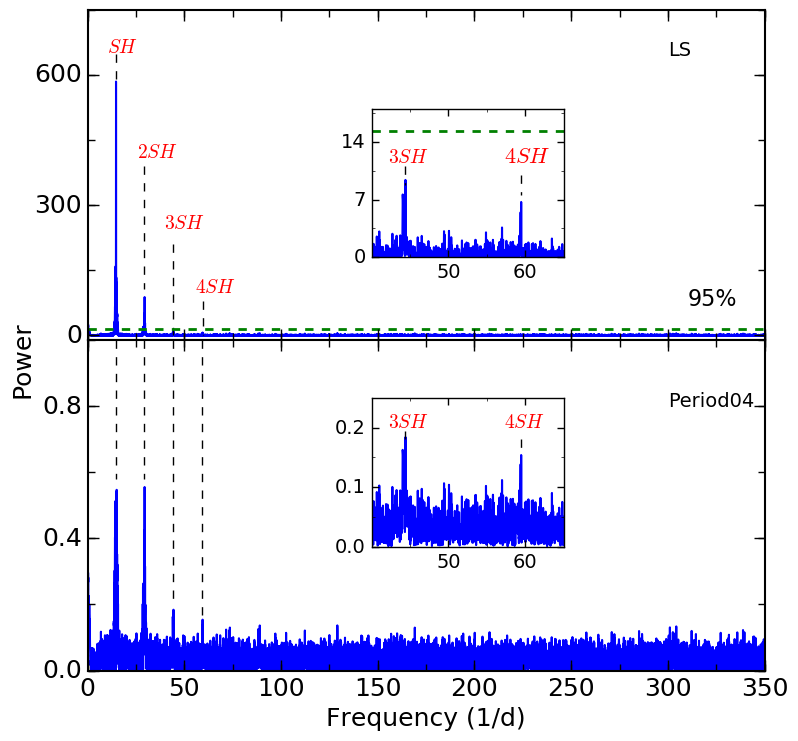}
\caption{LS and Period04 power spectra obtained from the combined detrended TESS light curve of J1100. The significant signals corresponding to the superhump frequency (SH) and its harmonics are distinctly marked. The significant frequencies observed in the power spectrum lie above the 95\% confidence level, which is represented by the dashed horizontal green line.}
\label{fig:tesslccyclesps_J1100}
\end{figure}

\begin{table*}
\caption{Periods corresponding to the dominant peaks in the power
spectra of J1100, J0935, and  J130559, as determined using Period04.\label{tab:ps}}
\setlength{\tabcolsep}{0.12in}
\centering
\begin{tabular}{@{}cccccc@{}}
\hline
\multicolumn{6}{c}{J1100} \\ 
\hline
Identification & \multicolumn{4}{c}{Period (d)} & \\ 
\cline{2-5} 
\hline
$P_{\rm SH}$   & 0.06786(1)$^\dagger$  & \\ 
\end{tabular}
\begin{tabular}{@{}ccc@{\hskip 1.0cm}ccc@{}}
\hline
\multicolumn{2}{c}{J0935} && \multicolumn{3}{c}{J130559} \\
\cline{1-2}\cline{4-6}
Identification & Period (d)  &&  Identification  & sector64-Period (d) & sector65-Period (d)  \\
\hline     
$P_{\Omega_{\text{pr}}^{\text{+}}}$  & 2.36(2)  &&   $P_{\Omega_{\text{pr}}^{\text{-}}}$  & 3.93(1)   & 3.73(1) \\
$P_{\rm SH_{+}}$  & 0.06584(2)  &&   $P_{\Omega}$      & 0.15092(2)   & 0.15092(1) \\
$P_{\Omega}$      & 0.06406(1)  &&   $P_{\rm SH_{-}}$  & 0.14519(3)   & 0.14516(4) \\
\hline
\end{tabular}
\\~\\
{$\dagger$ represents the periods derived from the combined detrended light curve of the \textit {TESS} data.}
\end{table*}


\section{Observations and data reduction}\label{sec:obs}
\label{sec:obs_tess}
A detailed log of the TESS observations for all three sources is given in Table \ref{tab:obslog}. The cadence for all three sources was 2 min.  The data of J1100, J0935, and J130559 are stored in the Mikulski Archive for Space Telescopes (MAST) data archive\footnote{\url{https://mast.stsci.edu/portal/Mashup/Clients/Mast/Portal.html}}.  TESS consists of four cameras, each with a field of view of 24$\times$24 degree$^2$, which are aligned to cover 24$\times$96-degree$^2$ strips of the sky called `sectors' \citep[see][for details]{Ricker15}. The TESS bandpass extends from 600 to 1000 nm, with an effective wavelength of 800 nm. Data taken during an anomalous event had quality flags greater than 0 in the FITS file data structure, and thus we have considered only the data with the `QUALITY flag' = 0. 
Simple aperture photometry (SAP) fluxes were used for analyzing all three sources.


\section{Analysis and results}\label{sec:analysis}


\subsection{CRTS J110014.7+131552}
 The TESS light curve of J1100 is shown in the top panel of Figure \ref{fig:tesslc_J1100}. During roughly 25 days of TESS observations, the system initially remained in a quiescent state for approximately 8 d. This was followed by a prolonged outburst lasting about 14.5 d, after which the system appeared to return to a quiescent state. The brightness difference with respect to quiescence, and thus the outburst amplitude, amounts to 4.5 mag. No other outbursts were visible during the TESS observation period. The amplitude, duration, and overall shape of the outburst indicate a superoutburst \citep[see, e.g., the light curve of V1504 Cyg in][]{Osaki14}, during which the formation of positive superhumps is expected. Indeed, Figure \ref{fig:tesszoomlc_J1100} shows zoomed-in versions of the shaded regions for specific time spans highlighted in top panel of Figure \ref{fig:tesslc_J1100} during the outburst maximum and decline phases, where the evolution of the superhumps is evident. A significant change in the observed hump pattern and amplitude is visible in each light curve panel. Well-resolved prominent humps are evident at the peak of the outburst, persisting for approximately 19 h, after which they begin to decrease in amplitude and gradually disappear. A second hump also appears to be associated with this prominent hump, which is clearly visible in Figure \ref{fig:tesszoomlc_J1100}. 

To identify the presence of periodicities in the system, the TESS data were first detrended to remove long-term variations, such as the outburst profile. We applied a locally weighted regression \citep[LOESS;][]{Cleveland79} with a smoothing span of 0.01 to smooth the light curves. This was then subtracted from the original light curve, resulting in the removal of the outburst variation (see middle panel of Figure \ref{fig:tesslc_J1100}). We utilized the Lomb-Scargle (LS) periodogram \citep[][]{Lomb76, Scargle82} and 
 Period04 \citep{Lenz04} algorithms to analyze the detrended light curve in order to search for periodic signals. Two prominent peaks are observed at frequencies of $\sim$ 14.7 c/d and $\sim$ 29.5 c/d, corresponding to periods of 0.06786(1) d  ($\sim$ 1.629 h) and 0.033896(7) d ($\sim$ 0.813 h), respectively from the combined detrened light curve of the TESS data (see Figure \ref{fig:tesslccyclesps_J1100}). The uncertainties in these periods were determined by performing 100 iterations of Markov Chain Monte Carlo (MCMC) simulations within the Period04 package. The significance of these detected peaks is determined by calculating the false-alarm probability \citep[FAP;][]{Horne86}. Both detected periods are found to be significant and lie above the 95\% confidence level represented by the dashed horizontal green line. A corresponding analysis of only the quiescent portion of the data instead yields a noise-dominated power spectrum without any significant peaks. The transient nature of the two frequencies detected in the bright data, as well as the shape of the modulation, unambiguously reveals these as being related to the superhump period. The strongest signal at $\sim$ 0.0678 d in the outburst data and thus corresponds to the superhump period ($P_\mathrm{SH}$). The second strongest peak can be identified with the second harmonic ($P_\mathrm{2SH}$). The third and fourth harmonics ($P_{\mathrm 3SH}$ and $P_{\mathrm 4SH}$) are also detected in the power spectra but are not significant, lying below the 95\% confidence level. The superhump profiles, amplitudes, and periods vary significantly during the superoutburst phases. To investigate periodic variations in different phases of the outburst light curve, we determined the times of superhump maxima from BJD 2460272.165 to 2460283.743, using a quadratic polynomial fitting method. Later, we determined the cycle numbers of the observed superhump maxima by adopting an initial epoch and a period of $\sim$ 0.0678 d. Using these cycle numbers, we performed a linear fit against the observed times of maxima to obtain a refined initial epoch and period, given by 
   \begin{equation}
 T_{\rm max}= BJD~2460272.1668 (10)+0.067796 (9) \times E.
   \end{equation}
    Using this refined epoch and period, we calculated the expected (calculated) timings of maxima. The O-C values were then computed by comparing the observed maxima times with these calculated times, which is a typical method for studying periodic variations and trends in superhump signals.  The O-C values are given in Table \ref{tab:midpoint} and plotted in the bottom panel of Figure \ref{fig:tesslc_J1100}.  The observed O-C trends provide clear evidence of superhump period variation after BJD 2460279.3. This expected period transition was also supported by periodogram analysis (not shown). The period was found to be shorter during the later phases of the outburst.

\begin{figure}[h!]
\centering
\subfigure[]
{\includegraphics[width=0.48\textwidth]{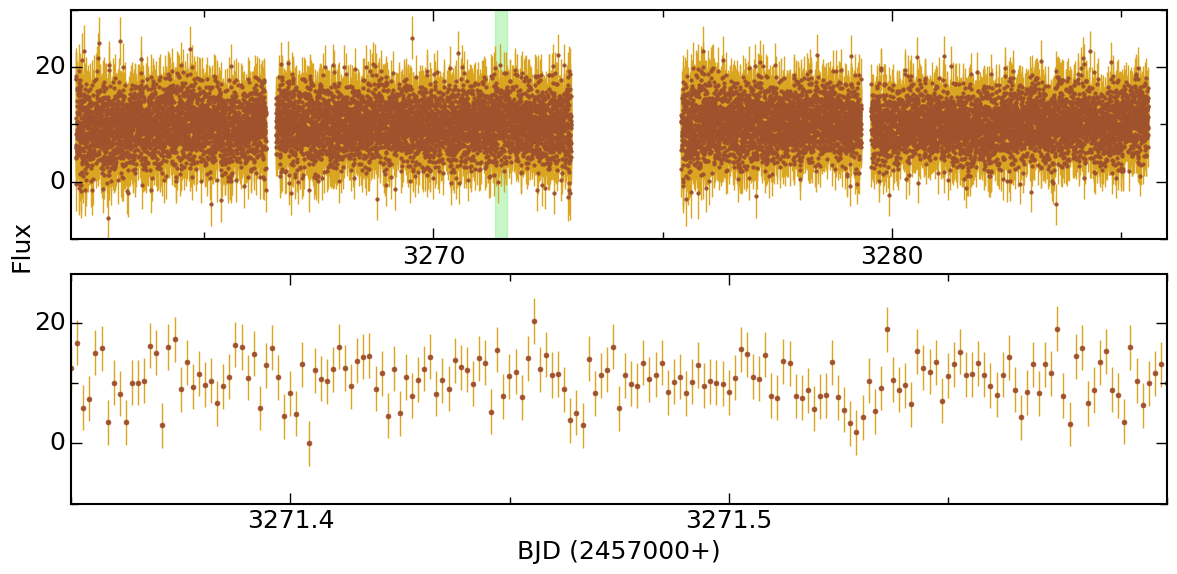}\label{fig:tesslc_J0935}}
\subfigure[]
{\includegraphics[width=0.5\textwidth]{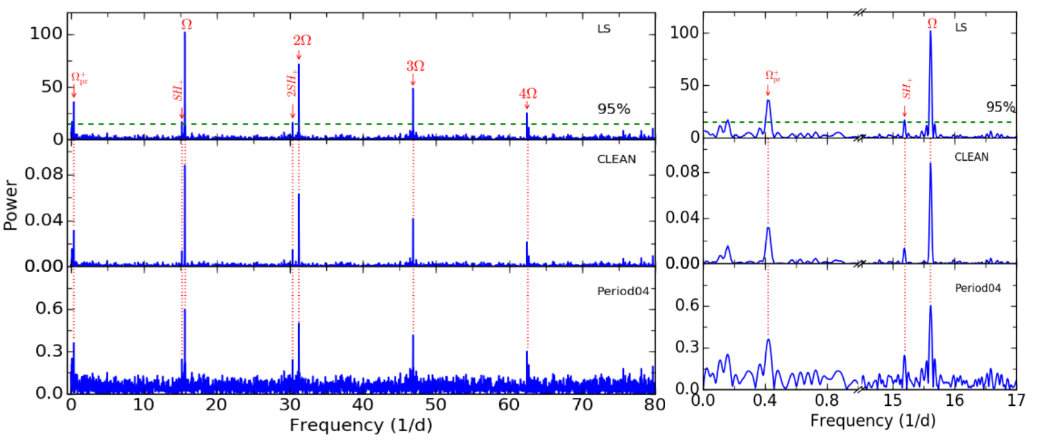}\label{fig:tessps_J0935}}
\caption{(a) Top panel: the full TESS light curve of J0935. Bottom panel: zoomed-in segment of the TESS light curve over $\sim$ 0.25 d, corresponding to the shaded region highlighted in the top panel, shown for clarity. (b) LS, CLEAN, and Period04 power spectra obtained from the combined TESS data. The significant signals are distinctly marked. A zoomed-in plot corresponding to the orbital ($\Omega$), positive superhump (SH${+}$), and prograde disc-precession ($\Omega_{\text{pr}}^{\text{+}}$) frequencies is also shown on the right.  All observed frequencies in the power spectrum lie above the 95\% confidence level, which is represented by the dashed horizontal green line. }
\end{figure}


\subsection{SDSS J093537.46+161950.8}
Figure \ref{fig:tesslc_J0935} presents the TESS observations of J0935, with the bottom subplot showing a zoomed-in view of the green-shaded region. Eclipse-like patterns are visible in the zoomed-in regions, though they do not appear very well-defined as compared to the eclipses reported by \cite{Southworth15}. Periodogram analyses using LS and Period04 were performed on the combined TESS data, which is shown in the top and bottom panels of Figure \ref{fig:tessps_J0935}, respectively, revealing several prominent frequencies. We calculated the FAP to evaluate the significance of the peaks. The detected periods that are found to be significant and lie above the 95\% confidence level are distinctly marked. Among those, the most prominent peak corresponds to 0.06406(1) d ($\sim$ 1.536 h), reflecting the orbital ($\Omega$) modulation \citep[see][]{Southworth15}. The corresponding harmonics up to 4$\times$$\Omega$ are also present.
 
 Next to the orbital period, we find a peak at 0.06584(2) d ($\sim$ 1.579 h), which is about 3\% longer than the orbital one. Such value is typical for positive superhumps ($P_{\rm SH_{+}}$).  The observed orbital and superhump periods are close, which may raise the possibility of misinterpretation due to flickering or marginal aliases. However, the detection of its harmonic, $P_{\rm 2SH_{+}}$, strengthens the interpretation of the signal as the superhump period.

At the low-frequency end, there is another strong peak at 2.36(2) d. In superhump terminology, the positive superhump period is interpreted as the beat period between the orbital period and the precession period ($P_{\Omega_{\text{pr}}^{\text{+}}}$) of an elliptical disc, i.e.,
$P_{\rm SH_{+}}$$^{-1}$ =$P_\Omega^{-1}$ $-$ $P_{\Omega_{\text{pr}}^{\text{+}}}^{-1}$, where $P_{\Omega_{\text{pr}}^{\text{+}}}^{-1}$ is the prograde disc-precession period when the disc precesses in the direction of the orbital motion \citep[apsidal precession;][]{Whitehurst88, Osaki89}.  Interpreting the 3\% longer period as $P_{\rm SH_{+}}$, we find $P_{\Omega_{\text{pr}}^{\text{+}}}$= 2.36 d, which agrees well with the observed low-frequency strong peak. One small peak also lies slightly above the significance level near the $P_{\Omega_{\text{pr}}^{+}}$ frequency region, but no combinations with the observed frequencies seem to be found. The periods corresponding to the orbital, positive superhump, and disc-precession signals are listed in Table \ref{tab:ps}, with their associated errors estimated using the MCMC method, as for J1100. To assess the reliability of the detected weak signals, we further performed a CLEAN power spectral analysis \citep{Roberts87}, which is shown in the middle panel of Figure \ref{fig:tessps_J0935}. This algorithm basically deconvolves window function from the dirty spectrum and produces a CLEAN spectrum, which is largely free of the many effects of spectral leakage. The CLEANed power spectrum  was obtained with a loop gain of 0.1 and 1000 iterations. All the orbital and superhump frequencies, along with their harmonics, are found to be present in the CLEANed power spectrum of J0935. Additionally, all these detected CLEANed periods are found to be consistent with the periods derived from the LS and Period04 power spectral analyses. This indicates that the marked periods derived from the extensive TESS observations are indeed real signals.  

\begin{figure}[h!]
\centering
\includegraphics[width=0.45\textwidth]{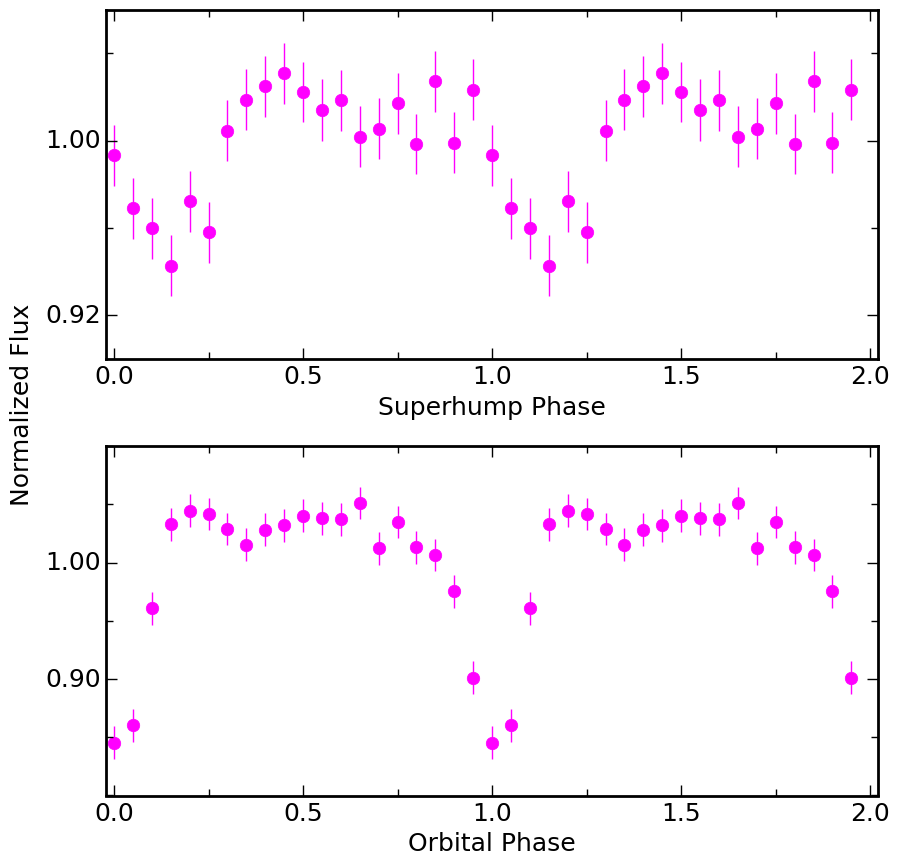}
\caption{Superhump-and-orbital phased light curves of J0935.} 
\label{fig:tessflc_ob_sh_J0935}
\end{figure}

To delve into periodic phenomena, the light curve was folded using  the zero-epoch provided by \cite{Southworth15} and our derived orbital and superhump periods. Figure \ref{fig:tessflc_ob_sh_J0935} shows the light curves folded at both the orbital and superhump frequencies. An eclipse feature is evident in the orbital-phased modulation, exhibiting symmetric ingress and egress profiles, along with a post-eclipse hump (see the bottom panel of Figure \ref{fig:tessflc_ob_sh_J0935}). On the other hand, the superhump modulation exhibits double-peaked modulation, with maxima occurring near phases 0.4 and 0.8, separated by a shallow minimum near phase 0.65 (see the upper panel of Figure \ref{fig:tessflc_ob_sh_J0935}). Evidence of a double-peaked pulse profile is also seen in the power spectrum, where a strong peak is observed at the superhump frequency as well as at its second harmonic.

\begin{figure}[h!]
\centering
\subfigure[]
{\includegraphics[width=0.48\textwidth]{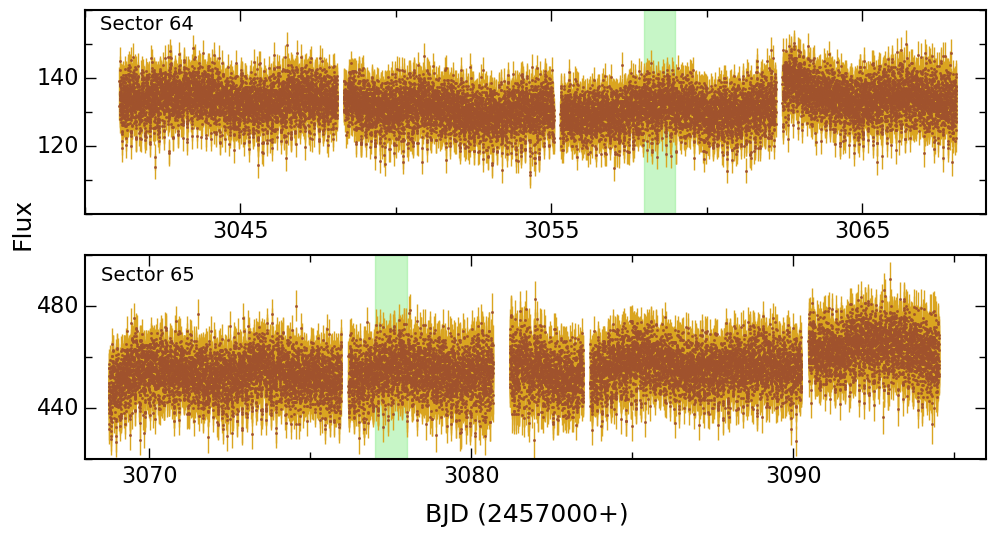}\label{fig:tesslc_H130559}}
\subfigure[]
{\includegraphics[width=0.48\textwidth]{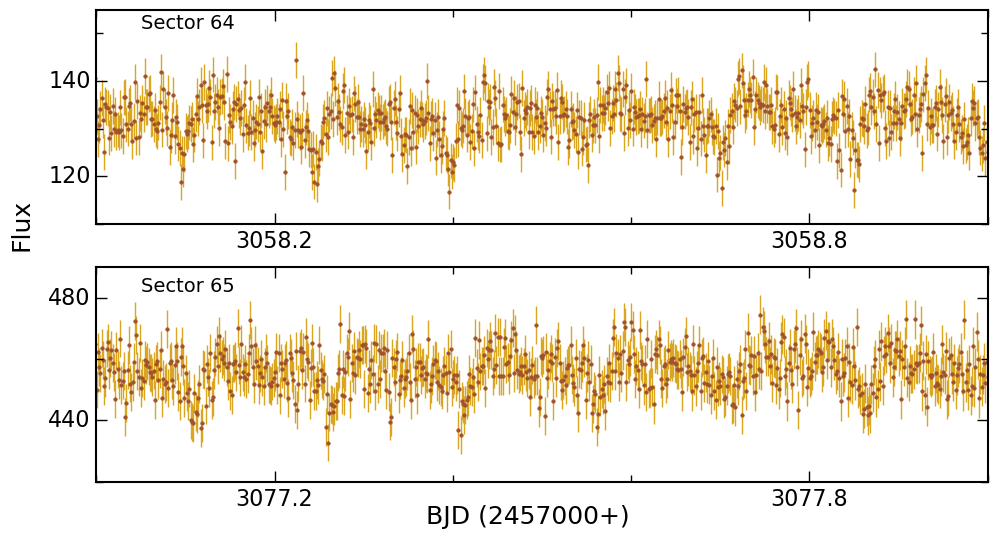}\label{fig:tesslc_zoomed_H130559}}
\caption{(a) The full TESS light curves of J130559 for sectors 64 and 65. (b) Zoomed-in segment of the TESS light curve over $\sim$ 1 d, corresponding to the shaded regions highlighted in Figure \ref{fig:tesslc_H130559}, shown for clarity.}
\end{figure}

 \subsection{[PK2008] HalphaJ130559} 
\begin{figure}[h!]
\centering
\subfigure[]
{\includegraphics[width=0.5\textwidth]{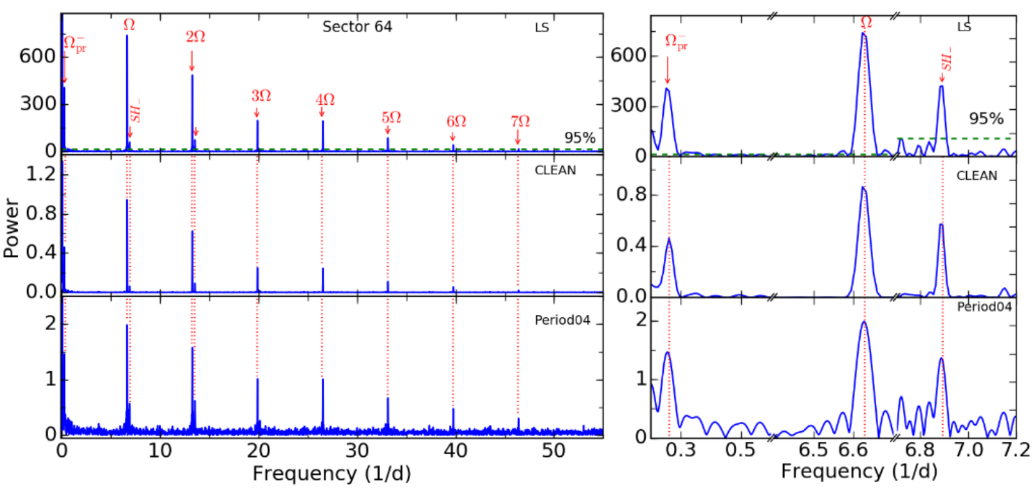}}
\subfigure[]
{\includegraphics[width=0.5\textwidth]{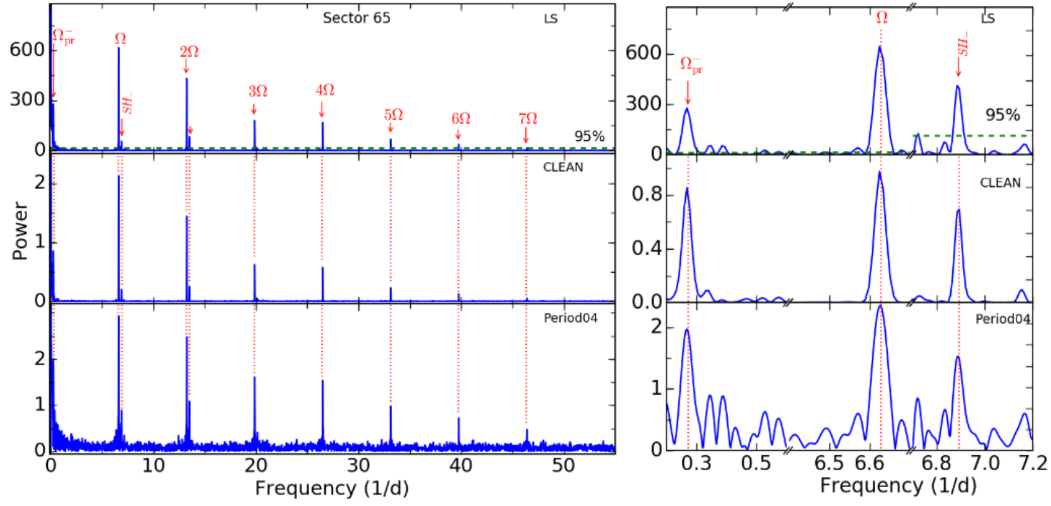}}
\caption{LS, CLEAN, and Period04 power spectra of J130559 obtained from the TESS data for (a) sector 64 and (b) sector 65. The significant signals are distinctly marked. A zoomed-in plot corresponding to the orbital ($\Omega$), negative superhump (SH$_{-}$), and  retrograde disc-precession ($\Omega_{\text{pr}}^{\text{-}}$) frequencies is also shown on the right. All observed frequencies in the power spectra lie above the 95\% confidence level, which is represented by the dashed horizontal green line.} 
\label{fig:tessps_H130559}
\end{figure}

Figures \ref{fig:tesslc_H130559} and \ref{fig:tesslc_zoomed_H130559} display the complete TESS light curves and the zoomed-in one-day light curves from the full observation period in sectors 64 and 65, respectively. The light curves from both sectors exhibit a clear eclipse profile, with each eclipse recurring approximately every 0.15 d, which thus should correspond to the orbital period of the system. To confirm this, the LS and Period04 periodogram algorithms were applied to the data from sectors 64 and 65, and the resulting power spectra, which show consistent peaks, are displayed in Figure  \ref{fig:tessps_H130559}. The power spectrum obtained for both sectors shows a dominant frequency corresponding to an average period of 0.15092(1) d  ($\sim$ 3.62 h), which is interpreted as the system's orbital period. The associated uncertainties were estimated using MCMC simulations implemented in the Period04 package. Higher harmonics corresponding to this frequency are also evident in each of the power spectra. Another period close to the orbital frequency is observed at an average period of   0.14517(3) d ($\sim$ 3.48 h), which differs from the orbital period by approximately 4\%. This could be associated with a negative superhump period ($P_{\rm SH_{-}}$). A significant peak corresponding to the period of $\sim$ 0.07398 d is also observed near the $2\Omega$ frequency in both sectors. However, this deviates slightly from its expected second harmonic period ($P_{\rm 2SH_{-}}$) of  $\sim$ 0.07259 d. These detected periods are found to be significant and lie above the 95\% confidence level. Consistent with the LS and Period04 results, the CLEAN algorithm also detected all significant peaks in the power spectra of both sectors  (see middle panels of Figure  \ref{fig:tessps_H130559}). The negative superhump period ($P_{\rm SH_{-}}$) is interpreted as $P_{\rm SH_{-}}$$^{-1}$ = $P_\Omega^{-1}$ $+$ $P_{\Omega_{\text{pr}}^{\text{-}}}^{-1}$, where $P_{\Omega_{\text{pr}}^{\text{-}}}^{-1}$ is the retrograde disc-precession period when the disc precesses in the opposite direction of the orbital motion or the node line precession \citep{Patterson97}. Using this expression and the derived orbital and candidate negative superhump periods, we calculated the disc-precession periods to be approximately 3.82 d for sector 64 and 3.80 d for sector 65. In the power spectra of both sectors, we observed peaks at periods of 3.93(1) d and 3.73(1) d in the low-frequency region for sectors 64 and 65, respectively (see Figure \ref{fig:tessps_H130559}).  
 An average value 3.83 d of these observed periods closely matches the expected values for sectors 64 and 65, although the individual sectors do not show this period precisely. Considering the available evidence, the identifications of the negative superhump and disc-precession periods remain provisional and subject to further confirmation.

\begin{figure*}[h!]
\centering
\subfigure[]
{\includegraphics[width=0.45\textwidth]{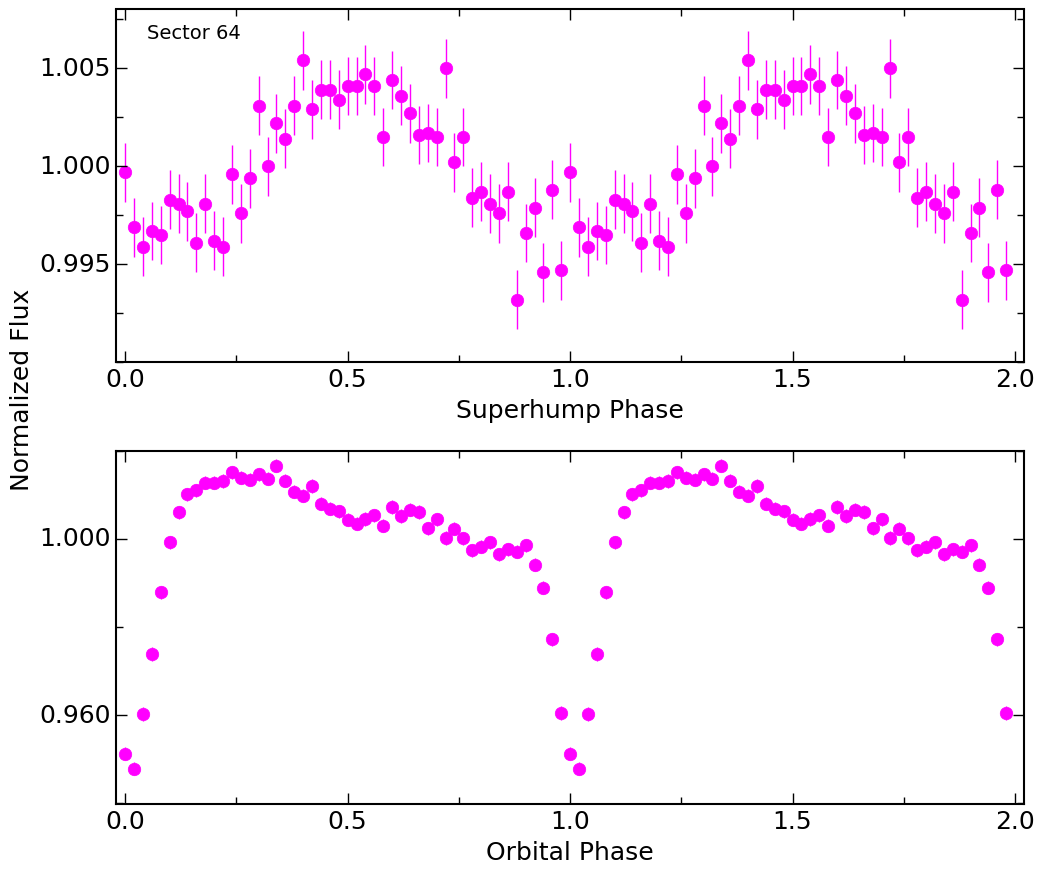}\label{fig:tessflcsec64_H130559}}
\subfigure[]
{\includegraphics[width=0.45\textwidth]{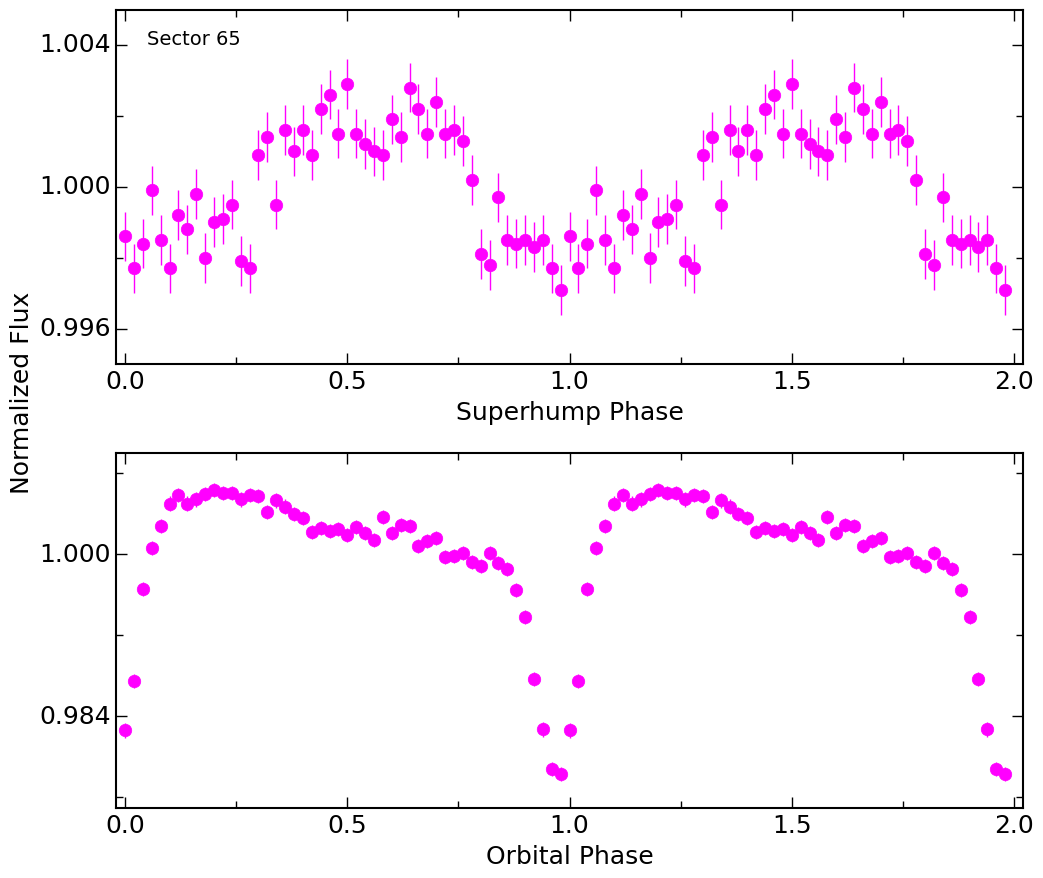}\label{fig:tessflcsec65_H130559}}
\caption{Superhump-and-orbital phase folded light curves of J130559 for (a) sector 64 and (b) sector 65.} 
\end{figure*}

Following the same approach as described for J1100 and J0935, we inspected the phased-light curve variations for J130559 using a random zero-epoch as BJD 2460042.0913 and our derived orbital and likely superhump periods, as shown in Figures \ref{fig:tessflcsec64_H130559} and \ref{fig:tessflcsec65_H130559} for sectors 64 and 65, respectively. A prominent eclipse feature is evident in the orbital phased-light curves of both sectors, which is associated with the post-eclipse hump. No noticeable hump is present at the phases before the primary eclipse. Apart from the primary eclipse, a secondary eclipse-like feature appears around phase 0.5. However, this feature seems to vanish in sector 65. The superhump-phased light curve displays a notable evolution between sectors 64 and 65. In sector 64, the superhump-phased light curve exhibits a dominant hump feature with a pronounced peak centered at phase 0.5. However, in sector 65, 
this peak undergoes a transformation, splitting into two separate peaks at phases 0.45 and 0.65, with a shallow dip appearing near phase 0.56. Finally, we note that a broad minimum is observed around phase 0.0 in the superhump-phased light curve.


\section{Discussion}\label{sec:diss}
\subsection{J1100}
TESS observations of J1100 reveal a prolonged outburst lasting 14.5 d following 8 d of quiescence, after which the system returns to a quiescent state. A precursor outburst is detected together with the main outbursts. Prominent superhumps emerged at the peak of the main outburst and took approximately one day to reach their maximum amplitude. As the outburst fades, the superhump amplitude progressively decreases and eventually disappears. This light curve feature is a typical characteristic of superoutbursts in SU UMa-type stars. 

A photometric period of 0.06786(1) d is detected from the combined TESS observations, which lies below the period gap and may correspond to the superhump period.
 This period is found to vary, becoming shorter during the later phases as the brightness decreases. The change in the superhump period can be attributed to variations in the disc radius throughout the superoutburst phases, supporting the thermal-tidal instability model for the superoutburst of SU UMa stars. The superhump period is typically longer during the early  phases of a superoutburst due to the expansion of the accretion disc, which reaches or exceeds the 3:1 resonance radius. At this stage, the disc becomes more eccentric, and the stronger tidal forces exerted by the secondary star cause the disc to precess more slowly, resulting in a longer superhump period. However, as the outburst progresses, the accretion disc gradually shrinks in size, which weakens the tidal interaction, allowing the disc to precess more quickly, which leads to a shorter superhump period in the later phases \citep{Kato09, Osaki13}. Additionally, the shape and amplitude of the light curve profile vary throughout the various phases of the outburst. During the early phase, the superhump modulation exhibits a sharply peaked pulse shape, which may arise from viscous dissipation in the periodically deforming disc. In the later phases, a second peak-like structure emerges, though it is not particularly prominent (see Figure \ref{fig:tesszoomlc_J1100}). If this feature indeed represents a secondary superhump maximum, it might be attributed to periodically variable dissipation at the accretion stream bright spot \citep[see][]{Wood11}. 


\subsection{J0935}
From the TESS observations, we have  detected an orbital period of 0.06406(1) d, with eclipsing features occurring periodically during its orbit, which is consistent with the orbital period determined by \cite{Southworth15}. A post-eclipse hump in orbital modulation is also observed, similar to that reported by \cite{Southworth15}. In many short-period CVs, a distinct orbital hump appears just before the eclipse, generally produced by the hot spot at the edge of the accretion disc rotating into view. On the other hand, the post-eclipse hump suggests the existence of a period different from the orbital period, known as the superhump period. Although only the orbital period was identified, no additional periods were observed in the results of \cite{Southworth15}, likely due to inadequate data coverage. In the present TESS observations, we found a significant peak at 0.06584(2) d, which is 3\% longer than the orbital period. This can be interpreted as a so-called common or positive superhump, the latter term referring to the positive period excess (i.e., the superhump period being longer than the orbital period). This likely arises from the prograde apsidal precession of an eccentric accretion disc. The observed disc-precession period of approximately 2.36 d is consistent with the interpretation of a positive superhump and supports the presence of a precessing eccentric disc in J0935. The superhump-phased light curve of J0935 shows a double-peaked structure separated by a shallow minimum. In past studies, most sources were found to exhibit normal pronounced bump-shaped superhump-phased light curves \citep[see][]{Patterson05, Bruch23a, Bruch23b, Bruch24}. Interestingly, some exceptions to these features have been found, as noted in the works of \citet{Schmidtobreick08} and \citet{Bruch23a, Bruch23b, Bruch24}, which identified peculiar sources with superhump light curves characterized by two or more peaks and sharply defined minima and maxima.  The fractional excess of the superhump period over the orbital period is expressed as 
\begin{equation}
\epsilon^{+} = \frac{P_{\rm SH_+} - P_{\Omega}}{P_{\Omega}}, 
\end{equation}
which is calculated to be approximately 0.028. The superhump excess has been shown to be correlated with the mass ratio of the binary system components by \cite{Patterson05} with the relation  
\begin{equation}
\epsilon^{+} = 0.18 \,q + 0.29 \,q^2,
\label{eq:epsilon}
\end{equation}
  although \cite{Kato22} suggests that this relation may suffer from the difficulty of not considering pressure effects within the accretion disc. The lack of outburst staging in J0935 prevents the application of stage-specific calibrations from \cite{Kato22}, and thus we rely on \cite{Patterson05} with caution. Taken at face value, Equation \ref{eq:epsilon} yields an estimate of $q$ approximately 0.128. Using this value and assuming the expected secondary mass corresponding to the orbital period of approximately  0.09\,$M_{\odot}$ \citep{Knigge11}, the mass of WD is estimated to be $\sim$0.7\,$M_{\odot}$ for J0935.

\begin{figure}
\centering
\includegraphics[width=0.48\textwidth]{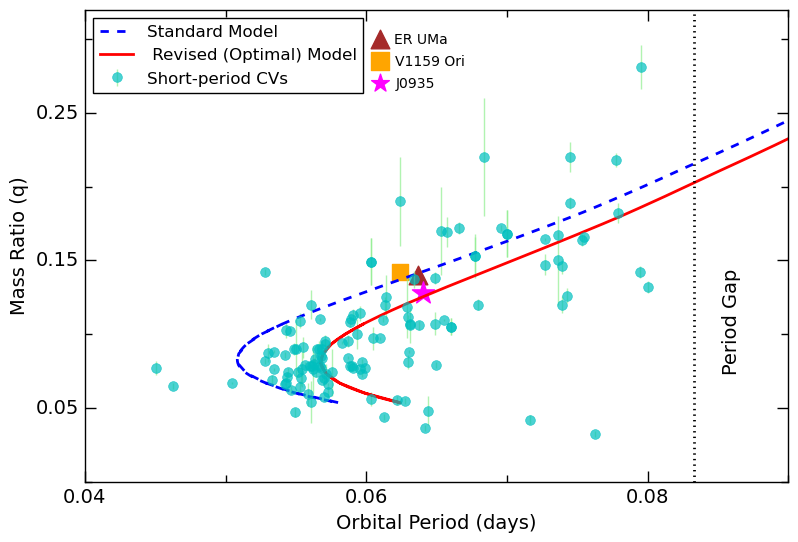}
\caption{The distribution of mass ratio versus orbital period for known short-period CVs. The dashed blue line shows the standard CV evolutionary
track from \cite{Knigge11}, while the solid red line represents their optimal binary track. The green points represent the mass ratio and period as reported in the literature \citep{Kato22}. V1159 Ori and ER UMa are represented by an orange square and a brown triangle, respectively, while J0935 (our work) is shown as a magenta star. } 
\label{fig:dist_q_Porb}
\end{figure}

For J0935,  we observe a short orbital period, a positive superhump period, a disc precession period, and its superhump period excess aligns with the \cite{Patterson98a} diagram. However, it has nonetheless remained in a constant average brightness state and has not exhibited any outburst activity during long-term observations, as discussed in the following paragraph.  In a few SU UMa-type systems, such as V1159 Ori, ER UMa, WZ Sge, EG Cnc, and V344 Lyr \citep{Patterson95, Patterson98, Gao99, Still10}, superhumps have also been detected outside the superoutburst intervals.
\cite{Hellier01} suggested that systems with $q\leq0.07$ can exhibit quiescent and normal outburst superhumps due to efficient angular momentum transfer by tidal dissipation in this regime, allowing both cooling and heating fronts to propagate even when the outer edge of the accretion disc extends beyond the 3:1 resonance radius. Moreover, the theoretical limit for the condition required to generate superhumps, or for the disc to reach the 3:1 resonance in dwarf novae, lies within the mass ratio range of
  $q$ = $0.22-0.39$ \citep{Whitehurst91,  Pearson06, Smak20}. The estimated mass ratio $q$ of 0.128 for J0935 does not meet either of the criteria required for superhumps. However, it is comparable to that of other superhumpers, such as V1159 Ori and ER UMa, where superhumps have been observed during quiescence or normal outburst \citep{Patterson95, Gao99}, with a mass ratio of approximately 0.14. The distribution of mass ratio versus orbital period for known short-period CVs including ER UMa, V1159 Ori, and J0935 is shown in Figure \ref{fig:dist_q_Porb}. The dashed blue line represents the standard CV evolutionary track from \cite{Knigge11}, while the solid red line illustrates their revised (optimal) binary track. The calculated mass ratio of J0935 falls within the predicted evolutionary track, implying that J0935 is following the typical evolutionary path for CVs. This system is currently in the middle stage of its evolution and is slowly transitioning toward a shorter orbital period. This process is expected to take a considerable amount of time before reaching its period minimum. Positive superhumps for CVs below the period gap are typically a short-term phenomenon intimately connected to superoutbursts. Within the TTI model, the deformation of the disc actually helps to deplete the disc (it enhances the mass-transfer through the disc). So in J0935, to keep that state and the positive superhumps, the donor star has to maintain a high mass-transfer rate to the disc. 

The discovery spectrum of J0935 features a prominent blue continuum and broad but weak Balmer emission lines, providing initial evidence that the system may be a nova-like variable with a high accretion rate \citep{Szkody09}. Along with the TESS, we also examined its long-term behavior in \textit{Zwicky} Transient Facility \citep[ZTF;][]{Bellm19, Graham19, Dekany20}, Catalina Real-Time Transient Survey \citep[CRTS;][]{Drake09}, American Association of Variable Star Observers \citep[AAVSO;][]{Kafka21}, and Global Astrometric Interferometer for Astrophysics \citep[{\em Gaia};][]{Gaiacollaboration_gaia_16, Gaiacollaboration_DR3_23} data, and did not find any evidence for outbursts. Some brightening measurements appear in a few epochs of the {\em Gaia} observations, but we suspect them to be spurious, as all G, BP, and RP magnitudes for those epochs were taken at the same time, yet their values are not aligned. Moreover, no such event was observed in the CRTS, even at or very close to the epochs of {\em Gaia} brightening. In addition, AAVSO data (although very incomplete) provides coverage of the object from May 2012 onward, with the only positive detections finding the object between V = 19.8 mag and 19.1 mag, while \cite{Szkody09} report g = 19.10 mag and u-g = 0.42 mag. The light curve from 
   \cite{Southworth15} shows the system at B = 19.2-18.6 mag out of eclipse, which, considering above positive blue colour index, lies within the same regime as the other brightness values. In summary, all measurements show the system at a visual brightness slightly fainter than 19 mag. While absence of evidence does not signify evidence of absence, there is no observational indication that the object has ever been in outburst. Therefore, it appears reasonable to assume that J0935 contains a stable accretion disc with the only significant brightness variations being related to orbital modulations such as eclipses, the orbital hump, and superhumps.

Adopting the Gaia DR3 distance of $1530^{+500}_{-280}$ pc from \cite{Bailer-Jones21} and Gaia magnitude of 18.73, we have estimated an absolute magnitude of $7.81^{+0.44}_{-0.62}$ and a color of 0.28(06) mag for J0935. These values place J0935 in a sparsely populated region of the colour–magnitude diagram presented by \citet[][their fig.\,2]{Abril20}, lying somewhat between the main bulks of all the CV subtypes, which makes it difficult to assign J0935 to any of them, based on the photometric characteristic. Its location on the diagram implies a mass-transfer rate that is neither particularly high, as in nova-likes, nor particularly low, as in dwarf novae below the period gap. Additionally, a comparison with the absolute magnitude versus orbital period plot in \cite{Mukai23}, which is mainly for intermediate polars (IPs) but also includes curves for dwarf novae in quiescence and outburst, is similarly inconclusive, as J0935 again lies between the two distributions. However, since J0935 is eclipsing, its high inclination will affect the absolute magnitude, causing it to appear about 3$-$4 mag fainter than at an average inclination \citep[see][]{Warner87, Selvelli19}. Accounting for this inclination effect places J0935 in the nova-likes/old novae/IPs regime in \cite{Abril20}, and in the high mass-transfer (high-luminosity IPs and dwarf novae in outburst) zones in \cite{Mukai23}.

As the SDSS spectrum of J0935 reveals strong He II ($\lambda$ 4686 \AA) emission, it may indicate that it belongs to the IP class of magnetic CVs. Interestingly, some IPs have been observed to show positive superhumps \citep[see][]{Mukai23}. However, high mass-transfer nova-likes have also been found to exhibit comparatively strong He II emission \citep[e.g.][]{Schmidtobreick15}, and hence, this does not appear to be an unambiguous indicator for a magnetic WD. Furthermore, the vast majority of IPs with orbital periods close to that of J0935 have a significantly lower intrinsic brightness, although we note that, with IGR J18173-2509, there is one example of a high-luminosity IP located below the period gap \citep{Mukai23}. Since this source has not been observed in X-rays so far, we do not know whether or not there are X-ray spin or sideband modulations that are typically used to confirm the IP nature of a system. Future X-ray observations may help to determine whether such modulations are present, and thereby confirm (or rule out) its classification as an IP.
   
The accumulated evidence suggests that J0935 is not an SU UMa-type dwarf nova, but instead may represent a rare type of high mass-transfer CV below the period gap, which makes it a particularly intriguing object. Based on the estimated mass ratio and orbital period, the system falls within the typical evolutionary track (see Figure \ref{fig:dist_q_Porb}). According to the standard theory of CV evolution, systems below the period gap are expected to have low mass-transfer rates and to exhibit dwarf nova outbursts. However, the unusually high mass-transfer rate suspected in this system deviates from these expectations, challenging the traditional understanding of CV evolution as provided by conventional models. The high mass-transfer rate in this short-period CV may be attributed to nova eruptions, as discussed by \cite{Patterson13} in their study of BK Lyn. Later, \cite{Hillman20} extensively discussed the effect of nova eruptions on the mass-transfer rate in the context of short- and long period CVs. As a result of these eruptions, the mass-transfer rate could remain high for extended periods. Over time, it might transform into a dwarf nova star, once the accretion rate has dropped sufficiently to trigger dwarf nova eruptions.


\subsection{J130559}
The orbital modulation of J130559 exhibits an eclipse feature characterized by asymmetric ingress and egress profiles, and a noticeable post-eclipse hump. A photometric period of 0.15092(1) d ($\sim$ 3.622 h) has been detected, which is interpreted as the orbital period of the system. This value differs significantly from the period of 0.16366(55) d ($\sim$ 3.928 h) obtained by \cite{Pretorius08} from time-resolved spectroscopy. The discrepancy is easily explained by the periodogram of the spectroscopic data (the bottom right plot in their fig.\,3) that presents a number of tightly spaced alias peaks, with four to five neighbouring frequencies exhibiting a power within $\sim$5 per cent of the strength of the one identified as the main peak.  In fact, the authors themselves emphasize that problem and point out the two closest peaks corresponding to periods of 4.273 and 3.635 h. The spectroscopic observations were taken on two nights separated by two days, which introduces alias peaks in the periodogram spaced by approximately 0.0208 c/h. Our measured frequency (0.2761 c/h, from $\sim$ 3.622 h) lies within 0.001 c/h of their 3.635 h peak (0.2751 c/h), but differs by $\sim$ 0.042 c/h from their reported 4.273 h value (0.2340 c/h), consistent with a $\pm$2-d alias. This indicates that 4.273 h is an alias introduced by their observational window. Notably, within the errors, their 3.635 h period agrees very well with the value derived from our eclipse data.

A significant period of 0.14517(3) d is also detected in the power spectrum of J130559, which is 4\% shorter than the orbital period. We tentatively interpret the modulation corresponding to this period as a negative superhump. The precession period is then $P_{\Omega_{\text{pr}}^{\text{-}}}^{-1}$ = $P_{\rm SH_-}^{-1}$ $-$ $P_\Omega^{-1}$.  In the TESS data from sectors 64 and 65, we find an average period of approximately 3.83 d that closely matches the expected disc-precession period. The occurrence of negative superhumps in CVs, is believed to result from the interaction between the secondary stream and the retrograde precession of a tilted disc \citep{Bonnet-Bidaud85, Harvey95, Wood09}. Being based on purely photometric evidence, the presence of a tilted disc in J130559 remains provisional. Additional phase-resolved spectroscopy and related analysis techniques such as Doppler tomography might have the potential to clarify the origin of such signals, but, to our knowledge, have not yet been applied to the case of tilted accretion discs in CVs.

The shape of suspected negative superhump-phased light curve of this system changes from a single prominent peak to an apparent separation into two peaks in sectors 64 and 65, respectively. Additionally, during the orbit, the secondary eclipse-like feature is visible in sector 64 and seems to disappear in sector 65. According to the current understanding of superhump mechanisms, it is expected that the the light distribution in the accretion disc and the location of the light source change over time. As a result, both the orbital and superhump waveforms are likely to be influenced by these variations. The two maxima during the superhump cycle can be attributed to the variable emission of energy caused by the matter stream from the secondary star overflowing the tilted accretion disc and impacting the disc at varying distances from the primary star. The change in the shape of superhump-phased light curve is also evident from the observed disc precession periods, which differ between sectors 64 and 65. This can likely be explained by variations in the location and brightness of light source over time. Such variations in the superhump waveforms have also been observed by \citep{Wood09} and \citep{Bruch24}, with the latter discussing how these variations generally vary with different disc precession phases. 

Using the negative superhump and orbital periods, the negative period excess is estimated as
\begin{equation}
\epsilon^{-} = \frac{P_{\rm SH_-} - P_{\Omega}}{P_{\Omega}}.
\end{equation}
 The  resulting values are $-0.0379\pm0.0002$ and $-0.0381\pm0.0003$ for sectors 64 and 65, respectively. To date, only 12 CVs are known to exhibit simultaneous negative superhumps and disc-precession periods \citep[see][]{Armstrong13}. Recently, LS Cam \citep{Rawat22}, SDSS J081256.85+191157.8 \citep{Sun23}, ASASSN-V J113750.23-572234.5 and ASASSN-17qj \citep{Sun24} have been added to this list. If the observed simultaneous periods are indeed negative superhump and disc-precession signals, then J130559 increases this number to seventeen, providing a new example for the study of tilted disc precession in CVs.    The most accepted model to explain the simultaneous presence of the negative superhump and disc-precession periods are the tilted and precessing (or 'wobbling') disc model, where lines of nodes of the accretion disc precess retrogradely. This causes a negative superhump period to arise due to the interaction between precession and orbital motions, although there is still no consensus on the exact physical process. In several studies, hydrodynamic simulations suggest that the tilted accretion disc responsible for negative superhumps may not be a rigid structure with a fixed inclination. Instead, it may behave as an unsteady, fluidic configuration that exhibits internal pressure-driven flows, variable vertical structure, and asymmetric or localized surface disturbances (‘bumps’) that can extend beyond the disc plane \citep{Montgomery09, Wood09, Thomas15}. These features could play a critical role in shaping the observed photometric modulations.

Using the relation between $\epsilon^{-}$ and $q$ given by \cite{Wood09} as 
\begin{equation}
q = -0.192 \left| \epsilon^{-} \right|^{1/2} + 10.37 \left| \epsilon^{-} \right| 
- 99.83 \left| \epsilon^{-} \right|^{3/2} + 451.1 \left| \epsilon^{-} \right|^2, 
\end{equation}
 the mass ratio is estimated to be approximately 0.267 and 0.270 for sectors 64 and 65, respectively.   These estimated mass ratios seem consistent with the improved mass ratio versus $\epsilon^{-}$ relation proposed by \cite{Stefanov23}, as shown in their figure 10b. With a secondary mass of 0.26 $M_\odot$ \citep{Knigge11} and an average mass ratio of approximately 0.27, the mass of the WD is estimated to be $\sim$ 0.96 $M_\odot$ for J130559.


\section{Summary}
\label{sec:sum}
The main conclusions of this study are summarized as follows:
\begin{enumerate}
\item A prolonged outburst associated with the precursor outburst is observed during the TESS observations of J1100, where prominent superhumps emerged during the maximum of the outburst with a period of approximately 0.06786 d. The superhump period is found to vary, becoming shorter during the later phases as the outbursts progress. The change in the superhump period can be attributed to variations in the disc radius throughout the superoutburst phases, supporting the thermal-tidal instability model for the superoutburst of SU UMa stars. Additionally, the amplitude and shape of the superhumps vary throughout the various phases of the outburst. During the early phases, the superhump signal seems to arise from viscous dissipation in the periodically deforming disc. However, in the later phases, a second peak-like structure appears, although it is not particularly prominent. If this feature is indeed a secondary superhump maximum, it could be due to the late stream-dominated source. 
 \item A disc-precession period of approximately 2.36 d along with a positive superhump period of $\sim$ 0.06584 d, which differs by 3\% from the orbital period, is observed in J0935. Such signals were not reported in earlier studies and may result from the prograde rotation of the deformed disc. In spite of superhumps and a short orbital period being mostly indicative of an SU UMa-type object, the complete lack of outburst activity, the characteristics of the discovery spectrum, and the absolute magnitude favor the interpretation that J0935 may be a high mass-transfer CV below the period gap, either a rare type of nova-like variable or a high-luminosity intermediate polar. This makes it an intriguing target for follow-up studies, because such objects are in contradiction to what is expected from our understanding of CV evolution.
\item An average periodicity of $\sim$ 3.83 d, along with periodicities of approximately 0.15092 d and 0.14517 d, are detected in J130559. The 0.15092 d signal confirms, for the first time, the orbital period of the system. The other two signals can be provisionally identified as the negative superhump and disc-precession periods. If the identified simultaneous signals truly reflect the negative superhump and disc-precession periods, then their origin may lie in the retrograde precession of a tilted disc and the interaction between the secondary stream and this precession, providing a new example for the study of tilted disc precession in CVs. The change in the shape of probable superhump light curves during different epochs can likely be explained by variations in the location and brightness of the light source over time.
\end{enumerate}



\begin{acknowledgements}
 We thank the anonymous referee for providing constructive comments and suggestions, which have significantly improved the manuscript. AJ and MC acknowledge support from the Centro de Astrofisica y Tecnologias Afines (CATA) fellowship via grant Agencia Nacional de Investigacion Desarrollo (ANID), BASAL FB210003. Additional support for MC is provided by ANID's FONDECYT Regular grant 1231637 and ANID's Millennium Science Initiative through grants ICN12\textunderscore 009 and AIM23-0001, awarded to the Millennium Institute of Astrophysics (MAS). CT acknowledges financial support by FONDECYT Regular Grant No 1211941. This paper includes data collected with the TESS mission, obtained from the MAST data archive at the Space Telescope Science Institute (STScI). Funding for the TESS mission is provided by the NASA Explorer Program. AJ acknowledges J. Joshi for reading our work and providing helpful comments. 
\end{acknowledgements}


\bibliographystyle{aa}
\bibliography{ref}


\begin{appendix}
\section{Additional Table}
   
\begin{table}[h!]
\centering
\caption{Observed times of maximum (BJD) and O-C values.}
\label{tab:midpoint}
\renewcommand{\arraystretch}{1.4}
\begin{tabular}{lccccccccc}
\hline
BJD (2457000+)& O-C & Error & BJD (2457000+) & O-C & Error & BJD (2457000+) & O-C & Error \\
\hline
 3272.165323  &   -0.001548    &   0.001268	  &  3277.392017  &    0.004847    &   0.002226         &     3280.305800  &    0.003399    &   0.004074      \\     
 3272.231329  &   -0.003339    &   0.000600       &  3277.458544  &    0.003578    &   0.002261         &     3280.369089  &    -0.001108   &   0.002437      \\
 3272.297471  &   -0.004993    &   0.000453       &  3277.524774  &    0.002012    &   0.001795         &     3280.441400  &    0.003407    &   0.006243      \\
 3272.363936  &   -0.006324    &   0.000399       &  3277.594595  &    0.004037    &   0.001962         &     3280.500826  &    -0.004964   &   0.002848      \\
 3272.432560  &   -0.005496    &   0.000334       &  3277.663931  &    0.005577    &   0.004341         &     3280.575513  &    0.001927    &   0.003873      \\
 3272.500037  &   -0.005815    &   0.000340       &  3277.730945  &    0.004795    &   0.003078         &     3280.644800  &    0.003418    &   0.003199      \\
 3272.567774  &   -0.005874    &   0.000311       &  3277.796907  &    0.002961    &   0.002764         &     3280.712600  &    0.003422    &   0.004574      \\
 3272.635732  &   -0.005712    &   0.000364       &  3277.862160  &    0.000418    &   0.002144         &     3280.778341  &    0.001367    &   0.006616      \\
 3272.703429  &   -0.005811    &   0.000415       &  3277.932527  &    0.002989    &   0.001774         &     3280.848191  &    0.003421    &   0.004798      \\
 3272.771835  &   -0.005201    &   0.000444       &  3278.000379  &    0.003044    &   0.002085	        &     3280.916000  &    0.003434    &   0.004264      \\
 3272.839471  &    -0.005361   &   0.000463       &  3278.071381  &    0.006250    &   0.002610      	&     3280.983800  &    0.003438    &   0.003452      \\
 3272.906289  &    -0.006339   &   0.000554       &  3278.134875  &    0.001948    &   0.002307         &     3281.049886  &    0.001728    &   0.003526      \\
 3275.485804  &    -0.003075   &   0.001471       &  3278.206521  &    0.005798    &   0.002623         &     3281.119028  &    0.003074    &   0.004774       \\
 3275.551078  &    -0.005598   &   0.001968       &  3278.276068  &    0.007549    &   0.004319         &     3281.187200  &    0.003450    &   0.003535       \\
 3275.621653  &    -0.002819   &   0.001740       &  3278.345400  &    0.009085    &   0.005706         &     3281.251481  &    -0.000065   &   0.002591       \\
 3275.687923  &    -0.004345   &   0.001818       &  3278.410356  &    0.006245    &   0.004503         &     3281.317862  &    -0.001481   &   0.002068       \\
 3275.756034  &    -0.004030   &   0.001776       &  3278.476535  &    0.004628    &   0.003411         &     3281.390600  &     0.003461   &   0.004252       \\
 3275.822708  &    -0.005152   &   0.001665       &  3278.548800  &    0.009097    &   0.004454         &     3281.458400  &     0.003465   &   0.004444       \\
 3275.894022  &    -0.001634   &   0.001612       &  3278.610982  &    0.003483    &   0.001983         &     3281.523499  &    0.000768    &   0.003858       \\     
 3275.957955  &    -0.005497   &   0.001824       &  3278.684400  &    0.009105    &   0.005041         &     3281.583574  &    -0.006953   &   0.002809        \\
 3276.027346  &    -0.003902   &   0.001537       &  3278.752200  &    0.009109    &   0.007344         &     3281.657707  &    -0.000616   &   0.004073        \\
 3276.097551  &    -0.001493   &   0.002101       &  3278.818791  &    0.007903    &   0.004296         &     3281.723199  &    -0.002920   &   0.002604        \\
 3276.168410  &     0.001570   &   0.002552       &  3278.880937  &    0.002253    &   0.002155         &     3281.797400  &    0.003485    &   0.017384        \\
 3276.231351  &    -0.003285   &   0.001577       &  3278.955600  &    0.009120    &   0.003459         &     3281.863005  &    0.001294    &   0.004374        \\
 3276.304990  &    0.002558    &   0.003781       &  3279.018877  &    0.004601    &   0.005230         &     3281.927536  &    -0.001971   &   0.001619        \\
 3276.367884  &    -0.002345   &   0.001690       &  3279.089893  &    0.007821    &   0.009547         &     3281.993985  &    -0.003318   &   0.002381        \\
 3276.438435  &    0.000410    &   0.001458       &  3279.159000  &    0.009132    &   0.004496         &     3282.068600  &    0.003501    &   0.006338        \\
 3276.502744  &    -0.003077   &   0.002065       &  3279.222498  &    0.004834    &   0.002032         &     3282.127515  &    -0.005381   &   0.002123       \\
 3276.573807  &    0.000190    &   0.001458       &  3279.289802  &    0.004342    &   0.002168         &     3282.196350  &    -0.004342   &   0.002122       \\
 3276.639897  &    -0.001516   &   0.001427       &  3279.553596  &    -0.003048   &   0.007053         &     3282.267932  &    -0.000556   &   0.002304       \\
 3276.706914  &    -0.002295   &   0.001623       &  3279.625956  &    0.001516    &   0.004336         &     3282.337107  &	  0.000823  &   0.002826       \\
 3276.775677  &    -0.001328   &   0.001921       &  3279.695600  &    0.003363    &   0.005266         &     3282.402122  &    -0.001958   &   0.003648        \\
 3276.848954  &    0.004153    &   0.004061       &  3279.763400  &    0.003367    &   0.005394         &     3282.469824  &    -0.002052   &   0.003155        \\
 3276.915951  &    0.003354    &   0.001926       &  3279.824671  &    -0.003158   &   0.003832         &     3282.538436  &    -0.001236   &   0.003267        \\
 3276.981992  &    0.001599    &   0.002601       &  3279.899000  &    0.003375    &   0.005249         &     3282.601745  &    -0.005723   &   0.002416        \\
 3277.048114  &    -0.000075   &   0.002634       &  3279.966800  &    0.003379    &   0.003587         &     3282.672352  &    -0.002912   &   0.002530        \\
 3277.119322  &    0.003337    &   0.003067       &  3280.029620  &    -0.001597   &   0.002240         &     3282.743979  &    0.000919    &   0.008072        \\
 3277.188214  &    0.004432    &   0.002625       &  3280.102400  &    0.003387    &   0.004853         &     3282.811493  &    0.000637    &   0.005441        \\
 3277.254823  &    0.003245    &   0.004440       &  3280.168574  &    0.001765    &   0.004877         &     3282.874878  &    -0.003774   &   0.004356         \\
 3277.323999  &    0.004625    &   0.003753       &  3280.233393  &    -0.001212   &   0.005377         &     3282.939223  &    -0.007225   &   0.005058        \\
 \hline                                                        \end{tabular}                                                 \end{table}                        
\clearpage
\begin{table}
\centering
\addtocounter{table}{-1}
\caption{(Continued)}
\label{tab:midpoint}
\renewcommand{\arraystretch}{1.4}
\begin{tabular}{lccc}
\hline
     BJD (2457000+) & O-C & Error \\
\hline
     3283.011525  &    -0.002720   &   0.003068   \\           
     3283.077412  &    -0.004629   &   0.002242   \\           3283.142773  &    -0.007064   &   0.002944   \\  
     3283.214729  &    -0.002904   &   0.004017   \\  
     3283.279408  &    -0.006021   &   0.004912   \\  
     3283.342114  &    -0.011111   &   0.007575   \\  
     3283.414145  &    -0.006876   &   0.002670   \\  
     3283.482390  &    -0.006427   &   0.002108   \\  
     3283.551882  &    -0.004731   &   0.002523   \\  
     3283.625551  &    0.001142    &   0.016449   \\  
     3283.680772  &    -0.011433   &   0.003915   \\  
     3283.743600  &    -0.016401   &   0.006551   \\       
\hline                                                         \end{tabular}                                                  \end{table}   

\end{appendix}
\end{document}